\documentclass{article}
\usepackage{arxiv}

\usepackage[utf8]{inputenc} % allow utf-8 input
\usepackage[T1]{fontenc}    % use 8-bit T1 fonts
\usepackage{hyperref}       % hyperlinks
\usepackage{url}            % simple URL typesetting
\usepackage{booktabs}       % professional-quality tables
\usepackage{amsfonts}       % blackboard math symbols
\usepackage{nicefrac}       % compact symbols for 1/2, etc.
\usepackage{microtype}      % microtypography
\usepackage{amsmath}
\usepackage{float}
\usepackage[caption=false]{subfig}
\usepackage{cleveref}       % smart cross-referencing
\usepackage{lipsum}         % Can be removed after putting your text content
\usepackage{graphicx}
\usepackage{doi}
\usepackage[font=small,labelfont=bf]{caption}

\title{Generative reconstruction of 3D volume elements for Ti-6Al-4V basketweave microstructure by optimization of CNN-based microstructural descriptors}

\author{ Vincent Bl\"umer \\
	\small Chair of Nonlinear Solid Mechanics\\
	\small University of Twente\\
	\small Enschede, The Netherlands \\
	\And
    Celal Soyarslan \\
	\small Chair of Nonlinear Solid Mechanics\\
    	\small Fraunhofer Innovation Platform\\
	\small University of Twente\\
	\small Enschede, The Netherlands \\
 	\And
    Ton van den Boogaard \\
	\small Chair of Nonlinear Solid Mechanics\\
	\small University of Twente\\
	\small Enschede, The Netherlands \\
}

\renewcommand{\shorttitle}{}

\begin{document}
\maketitle

\begin{abstract}
We present a methodology for the generative reconstruction of 3D Volume Elements (VE) for numerical multiscale analysis of Ti-6Al-4V processed by Additive Manufacturing (AM). The basketweave morphology, which is typically dominant in AM-processed Ti-6Al-4V, is analyzed in conventional Electron Backscatter Diffusion (EBSD) micrographs. Prior \(\beta\)-grain reconstruction is performed to obtain the out-of-plane orientation of the observed grains leveraging Burgers orientation relationship. Convolutional Neural Network (CNN) - based microstructure descriptors are extracted from the 2D data, and used for cross-section-based optimization of pixel values on orthogonal planes in 3D, using the Microstructure Characterization and Reconstruction (MCR) implementation MCRpy \cite{seibert2022microstructure}. In order to utilize MCRpy, which performs best for binary systems, the basketweave microstructure, which consists of up to twelve distinct grain orientations, is decomposed into several separate two-phase systems. Our reconstructions capture key characteristics of the titanium basketweave morphology and show qualitative resemblance to experimentally obtained 3D data. The preservation of volume fraction during assembly of the reconstruction remains an unadressed challenge at this stage. 
\end{abstract}

% keywords can be removed

\section{Introduction}
\label{introduction}

The accuracy of computed macroscale properties of any full-field multiscale analysis depends on the representativeness of the microscale solution domain. A volume element is representative, when it captures at least the minimum amount of microscale characteristics required to compute converged macroscale (or effective) properties. In other words, when the size of a Representative Volume Element (RVE) is increased, and therefore more information is captured, the computed macroscale properties do not change. For an extended discussion, see for example \cite{HILL1963,SAB1992,ShenBrinson2006,KanitForestJeulin20033647}. 

Computational multiscale studies of metals are usually based on EBSD micrographs, as they capture information about the material phase and crystal lattice orientation at the observed points in the micrograph domain. The single crystals are represented by elasticity with the applicable symmetry class and slip-system-based plasticity, a concept usually referred to as crystal plasticity. 

While 2D EBSD data can usually be acquired with reasonable effort, the generation of 3D data becomes tedious and often  infeasible. Unlike other techniques like Computer Tomography (CT), it cannot penetrate the material and needs to be performed on free surfaces. Therefore, the only way to generate 3D data is the sequential removal of material, surface preparation, and data recording. 3D EBSD studies that are especially noteworthy in this context are \cite{demott20203d,tsai2022development,chapman2021afrl}.

An alternative is the synthetic generation of RVEs by computational means and geometrical assumptions. This approach has long been common practice for example in steel microstructures that show a strong resemblance to Voronoi tessellations. The approximation error of the tessellation relative to the actual EBSD data is tolerated for the gain of 3D volumetric data. Due to the greatly reduced expense compared to the acquisition of 3D EBSD data through sequential slicing, it is a common choice for microstructures of low to moderate complexity. However, as the complexity of the microstructure increases, more details of the composition are lost.

The advent of Machine Learning (ML) in image recognition, sophisticated open-source modules for parallel optimization, and specialized computing hardware allows for new approaches to dimensionality expansion of microstructural data. In this contribution we develop a methodology for the generative reconstruction of 3D microstructural data for polycrystalline materials.

\section{Crystallography of Titanium}
As a basis for subsequent chapters, a brief introduction into the crystallographic properties of Titanium alloys is given.
\subsection{Phases and transformations}
At high temperatures, titanium generally occurs as \(\beta\)-titanium with body-centered-cubic (bcc) atomic lattice. At low/room temperature, \(\alpha\)-titanium with hexagonal-closest-packed (hcp) lattice is the dominant phase. Alloying elements can be categorized into either \(\alpha\)- or \(\beta\)-stabilizers, with Al being one of the former and V one of the latter. The alloy composition also influences the \(\beta\)-transus temperature. The widely used Ti-6Al-4V alloy transforms from \(\beta\) to \(\alpha\) phase during cooling at \(995^{\circ}\)C \cite{lutjering2007titanium}. It occurs in a wide range of room-temperature morphologies depending on processing history. 

The most commonly observed morphologies are globular (or primary) alpha, transformed \(\alpha+\beta\) (in the form of colonies or basketweave), bi-modal (when globular alpha is surrounded by transformed \(\alpha+\beta\)), and martensite \cite{pederson2002microstructure}. In wrought Ti-6Al-4V, bi-modal microstructure is typically observed. However, when processed by additive manufacturing, Ti-6Al-4V consists almost exclusively of transformed \(\alpha+\beta\). 

The formation of the transformed \(\alpha+\beta\) microstructure is governed by Burgers orientation relationship, which allows for twelve distinct variants of the transformation of the bcc- into the hcp-lattice. The emerging \(\alpha\)-grains have high aspect ratios and expand in so-called habit planes \cite{tong2017using}. In the formation of basketweave, all twelve variants of the \(\beta\) to \(\alpha\) transformation are active. In colonies, entire prior \(\beta\) grains transform along one of the variants, leading to a larger region of stacked \(\alpha\)-platelets with the same plate normal and crystal orientation \cite{makiewicz2012microstructure}. 

Titanium martensite can occur in the form of distorted-hcp \(\alpha'\)- and orthorhombic \(\alpha''\)-martensite. In Ti-6Al-4V, martensite creation is caused by the supersaturation of V in the hcp lattice. It should be noted that titanium martensite does not suffer from the characteristic loss of ductility as steel martensite, due to absence of carbon atoms. In the Ti-V system, a concentration of \(0-10\) at.\(\%\) V leads to the formation of \(\alpha'\), while a concentration of \(>10\) at.\(\%\) invokes \(\alpha''\) martensite \cite{mei2017martensitic}. The latter is not commonly observed in Ti-6Al-4V. 

\subsection{Microstructural evolution during the AM-process}
The Powder Bed Fusion (PBF) process is characterized by rapid, repeated heating and subsequent cooling. The molten powder solidifies as \(\beta\) in columnar grains that extend along the build direction. Upon cooling past the \(\beta\)-transus temperature, the \(\beta\)-titanium undergoes a martensitic transformation into \(\alpha'\)-martensite which is supersaturated with V and arranged into plates governed by Burgers orientation relationship. V is ejected to the boundaries of the \(\alpha'\) grains which decompose into \(\alpha+\beta\). \(\beta\) rods form on the boundary of the platelets \cite{tan2016revealing}. It is important to note that the transformed \(\alpha+\beta\) microstructure obtains its morphology from a martensitic transformation, but generally does not consist of martensite. Tan et al. \cite{tan2016revealing} observed martensite phase only in very thin printed samples.

Apart from a thin boundary layer where colonies are observed, the basketweave morphology is dominant. Formation of the basketweave microstructure can be attributed to the stabilizing temperature profile and coarsening \(\beta\)-grains \cite{makiewicz2012microstructure}. Higher cooling rates lead to finer basketweave and therefore increased hardness \cite{brandl2012microstructure}. 

AM-processed Ti-6Al-4V is characterized by large, columnar prior-\(\beta\)-grains, that extend along the print direction. Among these grains, the preferred orientation is the alignment of the \([001]_\beta\)-direction with the build direction. In the proximity of free surfaces of a printed part, prior-\(\beta\)-grains tend to be inclined towards the build direction at an angle and have distributed texture \cite{antonysamy2013effect}.

\section{Generative reconstruction of 3D data}
In the following sections, the necessary steps for the proposed methodology are presented. More detailed insights into Burgers orientation relationship as well as CNN-based microstructural descriptors are given.
\subsection{Acquired data}
Point of departure for the 3D reconstruction is a EBSD micrograph taken from a tensile test specimen manufactured by Laser Powder Bed Fusion (L-PBF). The physical dimensions of the micrograph are \(77.86 \,\mu\textup{m} \times 58.09\,\mu\textup{m} \) and the resolution is \(1072 \times 800\). 

\begin{figure}[H]
\centering
\includegraphics[width=.45\textwidth]{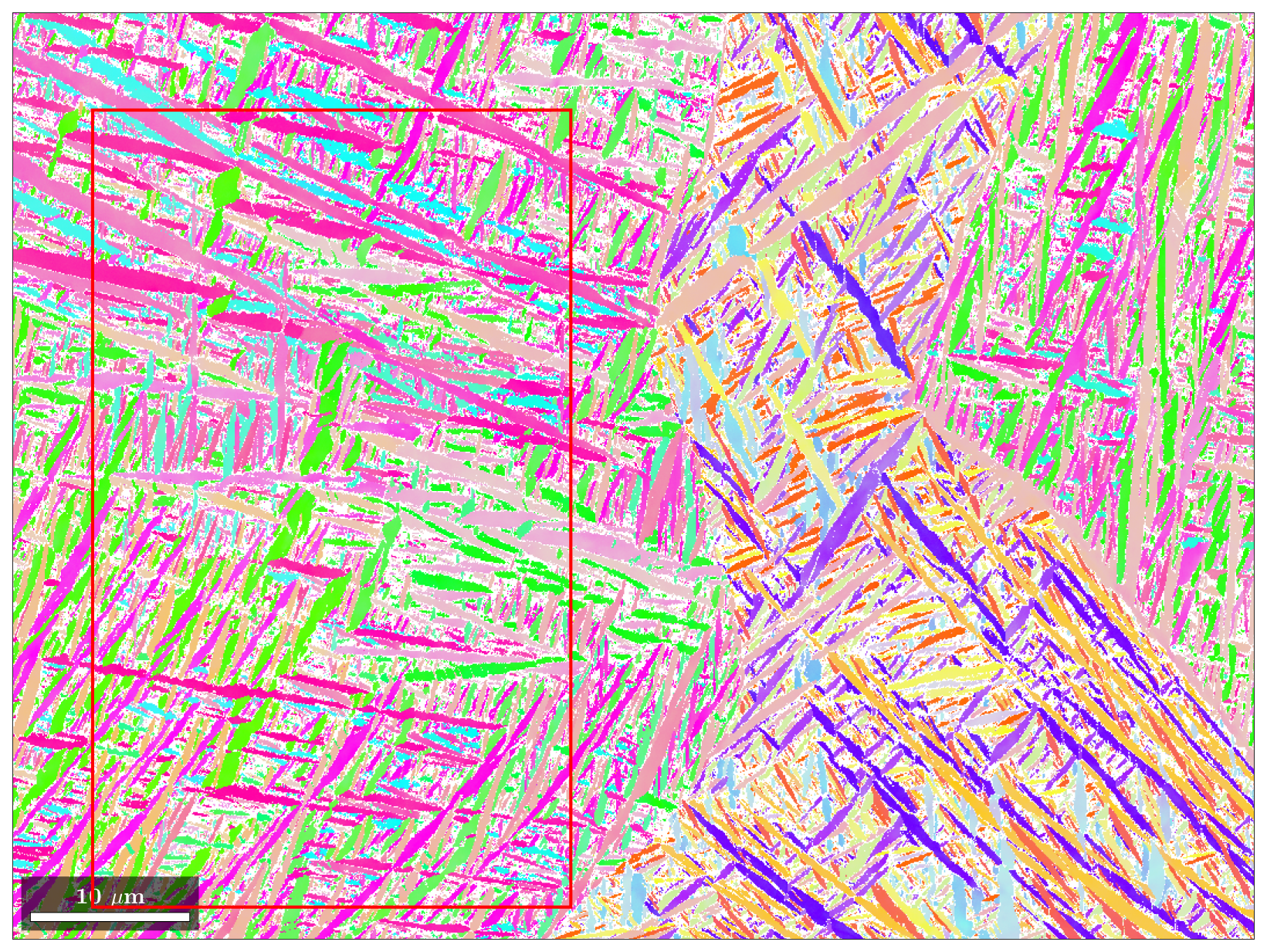}
\caption{EBSD micrograph of an additively manufactured tensile test specimen built from Ti-6Al-4V. The vertical axis is aligned with the build direction. The rectangular region is cropped out for further study.}\label{full}
\end{figure}

The build direction is aligned with the vertical axis of the image. Basketweave microstructure can be observed everywhere in the micrograph. The separation of the region into elongated parent grains is apparent to the eye. The alignment convention of the crystal reference frame is \(\boldsymbol{X}\parallel  \boldsymbol{a}^*,\,\, \boldsymbol{Y}\parallel  \boldsymbol{b},\,\,\boldsymbol{Z}\parallel  \boldsymbol{c}^*\), where \(\boldsymbol{X},\boldsymbol{Y},\boldsymbol{Z}\) are the orthogonal axes of the reference coordinate system and \(\boldsymbol{a},\boldsymbol{b},\boldsymbol{c}\) are the crystal axes of the hcp lattice, where \(\boldsymbol{a}\) and \(\boldsymbol{b}\) are inclined at a \(120^{\circ}\) angle. The asterisk denotes the reciprocal of an axis. They are computed as
\begin{equation}
    \boldsymbol{a}^*=2\pi\frac{\boldsymbol{b}\times\boldsymbol{c}}{\boldsymbol{a}\cdot\left ( \boldsymbol{b}\times 
\boldsymbol{c}\right )} 
,\quad
\boldsymbol{b}^*=2\pi\frac{\boldsymbol{c}\times\boldsymbol{a}}{\boldsymbol{a}\cdot\left ( \boldsymbol{b}\times \boldsymbol{c}\right )} 
,\quad
\boldsymbol{c}^*=2\pi\frac{\boldsymbol{a}\times\boldsymbol{b}}{\boldsymbol{a}\cdot\left ( \boldsymbol{b}\times \boldsymbol{c}\right )} 
\end{equation}
respectively.
\subsection{Parent grain reconstruction}
Parent grain reconstruction is performed using Matlab MTEX \cite{bachmann2010texture}. The reconstruction is based on Burgers orientation relationship, according to which each \(\alpha\)-grain should match one of twelve misorientations to the parent grain from which it emerged. The twelve variants for alignment are given in Table \ref{Table1}.
\begin{table}[H]
\centering
\caption{The twelve variants of Burgers orientation relationship \cite{furuhara1992crystallography}. Each row indicates one possible axis-alignment of the emerging \(\alpha\)-grain to the parent grains axes.
}
\scalebox{0.8}{\begin{tabular}{c c c c} 
Variant ID & \(\boldsymbol{Z}_\alpha\) & \(\boldsymbol{Y}_\alpha\) & \(\boldsymbol{X}_\alpha\) \\
\hline 
\\[-0.9em]

V1  & \((110)_\beta \parallel (0001)_\alpha\)              & \([\overline{1}1\overline{1}]_\beta\parallel[11\overline{2}0]_\alpha\)  & \([\overline{1}12]_\beta \parallel [\overline{1}100]_\alpha\) \\ 
V2  & \((110)_\beta \parallel (0001)_\alpha\)              & \([1\overline{1}\overline{1}]_\beta\parallel[11\overline{2}0]_\alpha\)  & \([\overline{1}1\overline{2}]_\beta \parallel [\overline{1}100]_\alpha\) \\ 
V3  & \((1\overline{1}0)_\beta \parallel (0001)_\alpha\)   & \([11\overline{1}]_\beta\parallel[11\overline{2}0]_\alpha\)             & \([112]_\beta \parallel [\overline{1}100]_\alpha\) \\ 
V4  & \((1\overline{1}0)_\beta \parallel (0001)_\alpha\)   & \([111]_\beta\parallel[11\overline{2}0]_\alpha\)                        & \([\overline{1}\overline{1}2]_\beta \parallel [\overline{1}100]_\alpha\) \\ 
V5  & \((011)_\beta \parallel (0001)_\alpha\)              & \([11\overline{1}]_\beta\parallel[11\overline{2}0]_\alpha\)             & \([\overline{2}1\overline{1}]_\beta \parallel [\overline{1}100]_\alpha\) \\ 
V6  & \((011)_\beta \parallel (0001)_\alpha\)              & \([1\overline{1}1]_\beta\parallel[11\overline{2}0]_\alpha\)             & \([21\overline{1}]_\beta \parallel [\overline{1}100]_\alpha\) \\ 
V7  & \((01\overline{1})_\beta \parallel (0001)_\alpha\)   & \([\overline{1}11]_\beta\parallel[11\overline{2}0]_\alpha\)             & \([211]_\beta \parallel [\overline{1}100]_\alpha\) \\ 
V8  & \((01\overline{1})_\beta \parallel (0001)_\alpha\)   & \([111]_\beta\parallel[11\overline{2}0]_\alpha\)                        & \([2\overline{1}\overline{1}]_\beta \parallel [\overline{1}100]_\alpha\) \\ 
V9  & \((101)_\beta \parallel (0001)_\alpha\)              & \([\overline{1}11]_\beta\parallel[11\overline{2}0]_\alpha\)             & \([\overline{1}\overline{2}1]_\beta \parallel [\overline{1}100]_\alpha\) \\ 
V10 & \((101)_\beta \parallel (0001)_\alpha\)              & \([11\overline{1}]_\beta\parallel[11\overline{2}0]_\alpha\)             & \([\overline{1}21]_\beta \parallel [\overline{1}100]_\alpha\) \\ 
V11 & \((\overline{1}01)_\beta \parallel (0001)_\alpha\)   & \([1\overline{1}1]_\beta\parallel[11\overline{2}0]_\alpha\)             & \([121]_\beta \parallel [\overline{1}100]_\alpha\) \\ 
V12 & \((\overline{1}01)_\beta \parallel (0001)_\alpha\)   & \([111]_\beta\parallel[11\overline{2}0]_\alpha\)                        & \([\overline{1}2\overline{1}]_\beta \parallel [\overline{1}100]_\alpha\) \\ 
\\[-0.9em]
\hline 
\end{tabular}}

\label{Table1}
\end{table}
For each \(\alpha\)-grain, its association to one parent-grain, as well as the matching variant is computed. The orientation of the parent grain is also reconstructed. 

\begin{figure}[H]
\centering
\includegraphics[width=.5\textwidth]{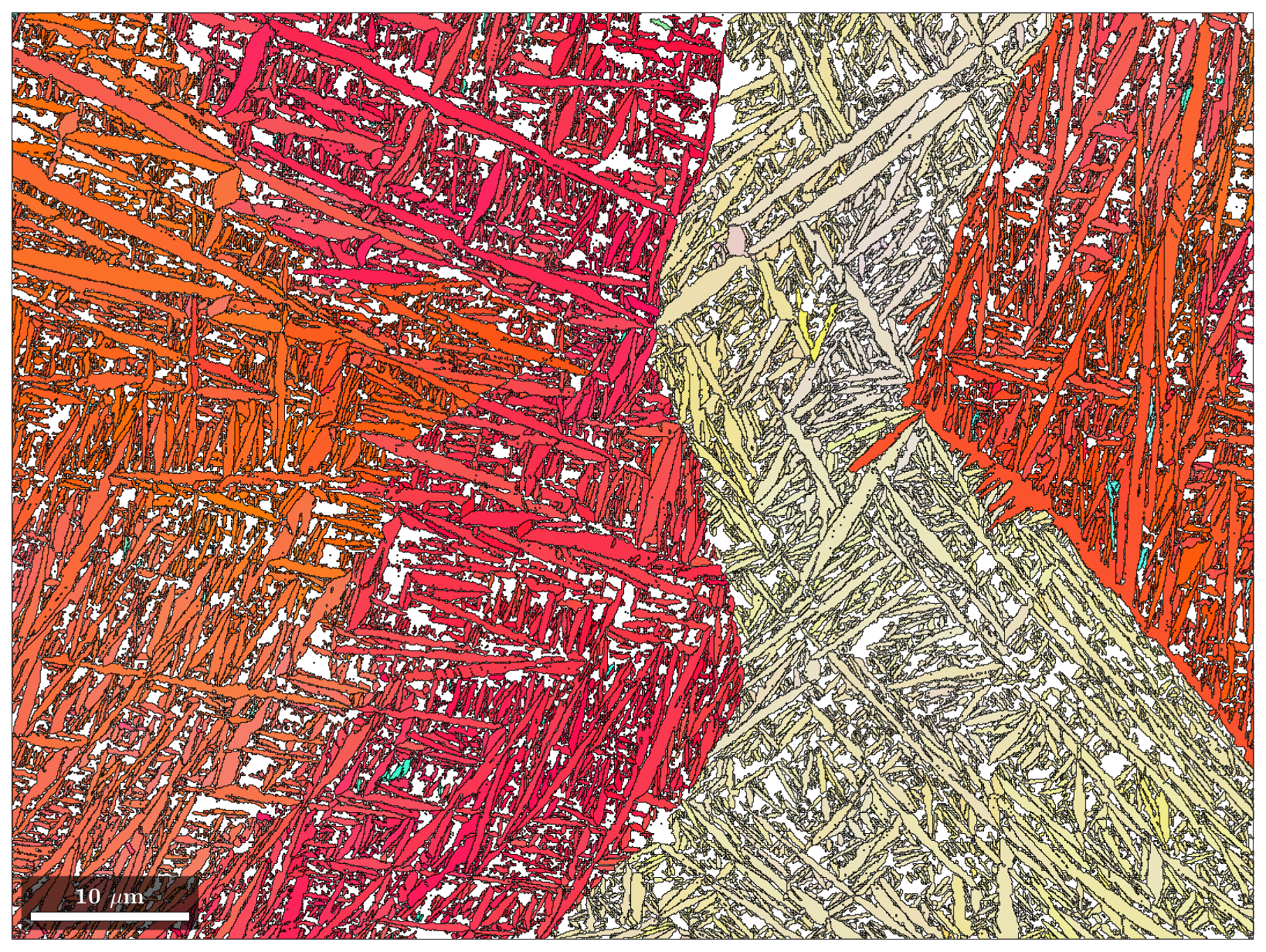}
\caption{Computed orientations of parent grains after prior-\(\beta\) grain reconstruction is performed in Matlab MTEX.}\label{parent_rec}
\end{figure}

Figure \ref{parent_rec} illustrates the association of the \(\alpha\)-grains to their respective parent grains and confirms the first visual impression. The rectangular region shown in Figure \ref{full} is used for further study. This is done to isolate the microstructural information of one parent grain from blending with the statistics of other regions of the EBSD.

\begin{figure}[H]
\centering
\includegraphics[width=.3\textwidth]{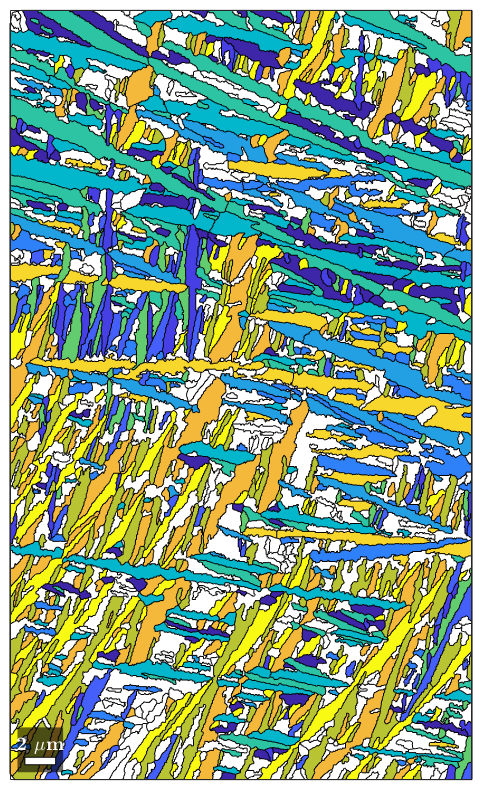}
\caption{During the prior-\(\beta\) grain reconstruction, each \(\alpha\) grain is matched to one of the twelve variants of Burgers orientation relationship.}\label{vars}
\end{figure}

Figure \ref{vars} shows the computed misorientation variants of the \(\alpha\) grains. It is evident that all grains that belong to one of the misorientation variants (hereafter called grain group) have almost identical grain axis orientations in the 2D observation plane (in addition to their identical crystal lattice orientations). The preferred growth direction of an \(\alpha\)-grain is called the habit plane, which is normal to one of the family of \([1100]_\alpha\)-directions of the \(\alpha\)-crystal after transformation \cite{tong2017using}. Therefore, the grain axis orientation in the out of plane dimension can be reconstructed up to the ambiguity caused by the rotational symmetry of the hcp lattice. 

\begin{figure}[H]
\centering
\includegraphics[width=.7 \textwidth]{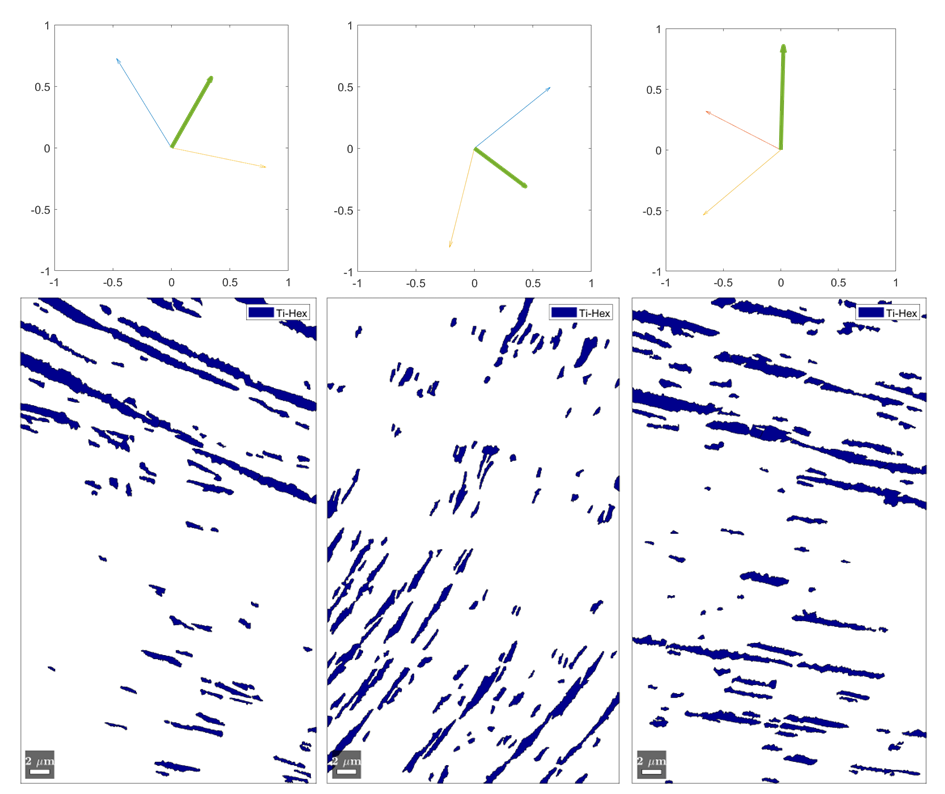}
\caption{All \(\alpha\)-grain groups are extracted from the EBSD and their habit plane normals are computed. This process is shown for three of the twelve groups.}\label{normals}
\end{figure}

Applying Burgers orientation relationship yields three vectors that potentially represent a grain group's habit plane normal. From these vectors, the one that is closest to in-plane orthogonal to the in-plane grain axis is selected as the most plausible habit plane normal. Results of this process are shown in Figure \ref{normals}. The habit plane normals are later used to assemble the microstructural descriptors in orientations consistent with the 3D orientations of grains.

\subsection{Microstructural Descriptor}
MCRpy is an open-source software platform used for characterization and reconstruction of microstructures. Reconstructions are obtained through optimization of pixel values in a reconstruction domain to minimize a loss function. The choice of descriptor is flexible and ranges from \(n\)-point statistics to more novel choices like Gram matrices \cite{gatys2016image}, which have lead to promising results on recent years \cite{li2018transfer}. Gram matrices are calculated from the activations of a pre-trained convolutional neural network when exposed to an image. MCRpy uses the general image recognition network VGG-19 \cite{simonyan2014very} with ImageNet weights as a reference \cite{seibert2022microstructure}. 
\subsection{Reconstruction of grain groups}
A characteristic property of titanium basketweave microstructure is that all \(\alpha\)-grains emerging from the same parent grain have one out of twelve distinct orientations. Not all of the continuous space of possible orientations is observed in the microstructure, which allows the representation of the microstructure as a discrete set of twelve phases. MCRpy is primarily built for the reconstruction of two-phase microstructures, by varying pixel values in a continuous range from 0 to 1. Although the applied concepts extend to higher number of phases, and multiphase reconstruction is implemented in MCRpy, the addition of a third phase already leads to a steep increase in computational time. A potential circumvention of this limitation is the separation of the microstructure into twelve separate reconstruction problems and subsequent superposition. Each of the grain groups is isolated from the EBSD and represented as phase 1 against a background of phase 0. Reconstructions are performed over \(100\times100\times100\) grid domains where the out-of-plane descriptors of the phases are assembled consistent with the orientation of the grain group's habit plane normal. A selection of results is shown in the following figures. For a visual comparison to experimentally obtained data, the reader is referred to \cite{demott20203d}.

\begin{figure}[H]
\centering
\subfloat[The 7\textsuperscript{th} grain group with a volume fraction of 8.35\%]{%
  \includegraphics[trim={0 2.5cm 0 1cm},clip,width=.7\columnwidth]{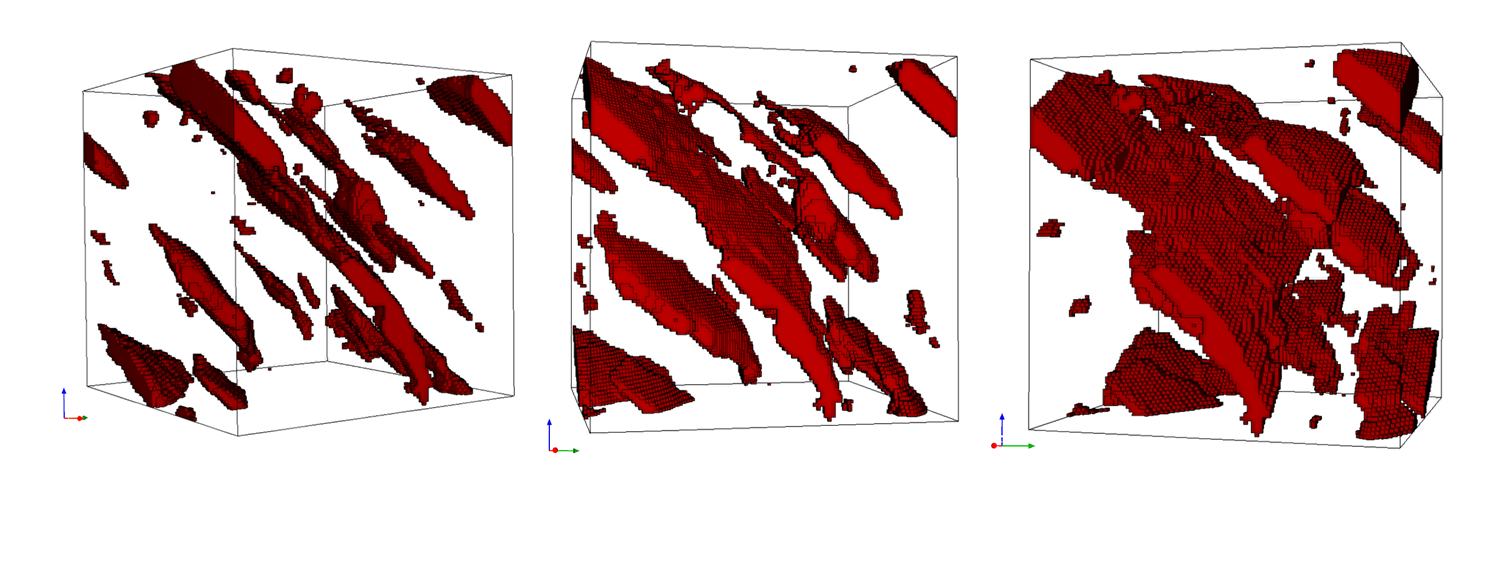}
}

\subfloat[The 12\textsuperscript{th} grain group with a volume fraction of 8.47\%]{%
  \includegraphics[trim={0 2.5cm 0 1cm},clip,width=.7\columnwidth]{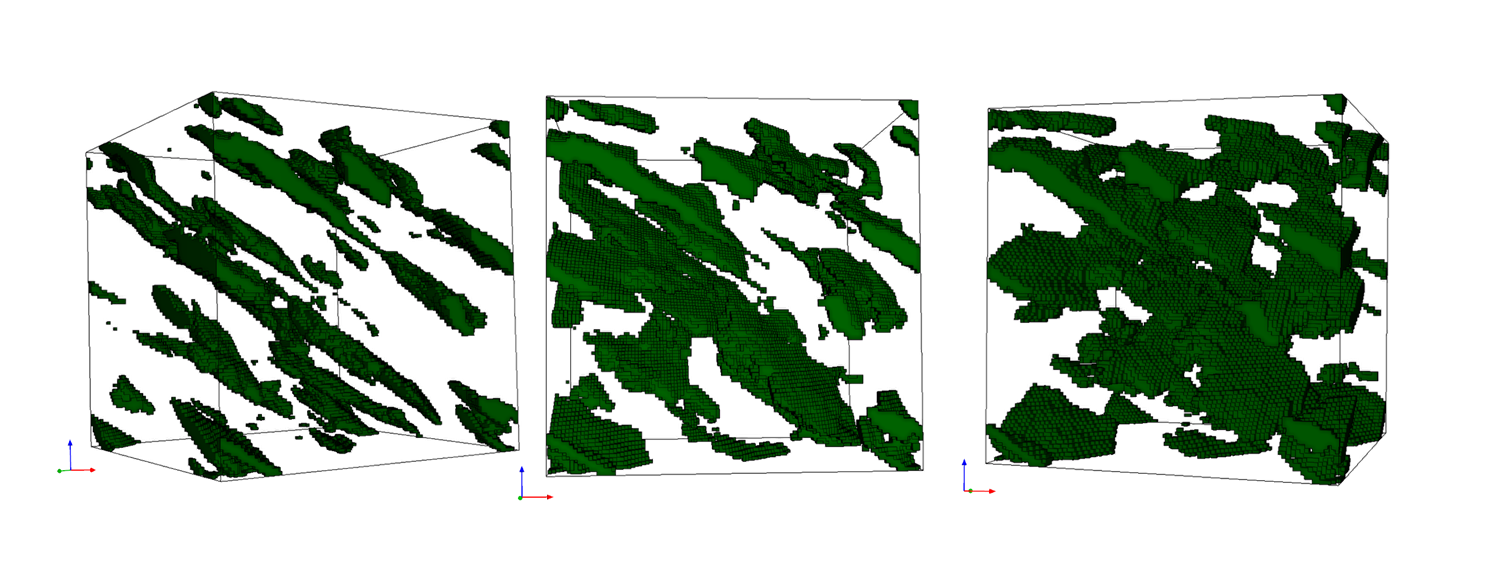}%
}

\subfloat[The 6\textsuperscript{th} grain group with a volume fraction of 10.34\%]{%
  \includegraphics[trim={0 2.5cm 0 1cm},clip,width=.7\columnwidth]{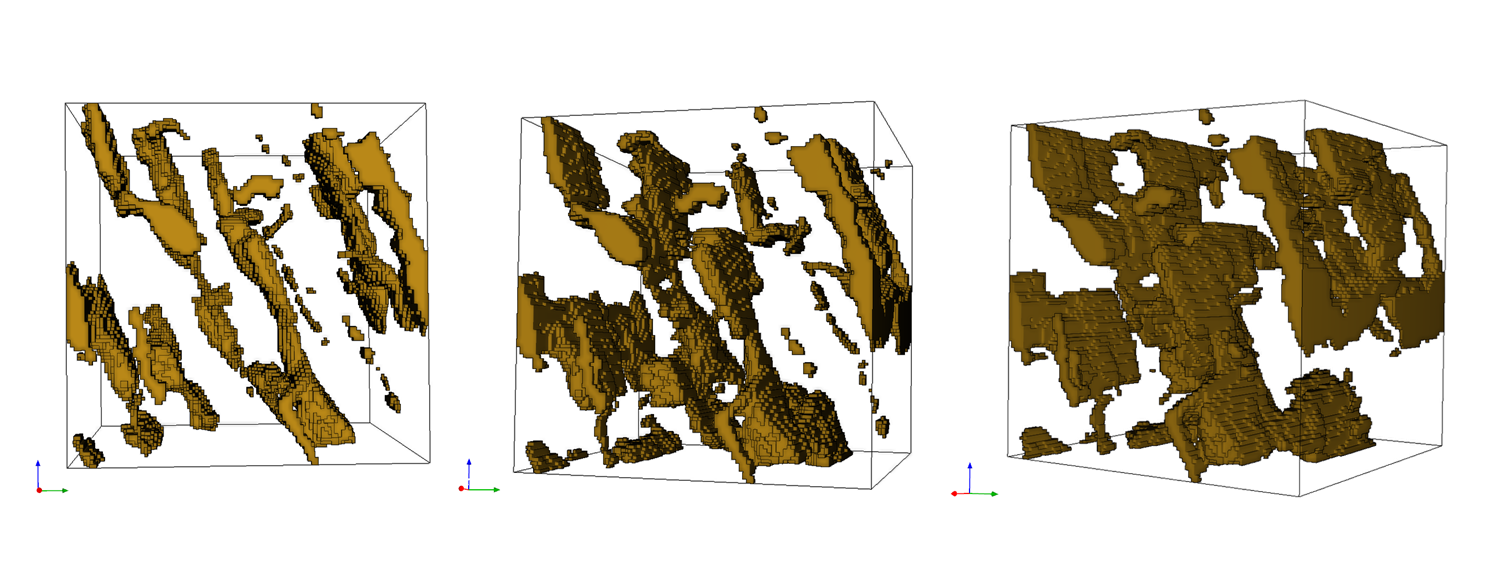}%
}

\caption{Three-dimensional reconstructions of three distinct groups from a set of twelve categorized grain types.}

\end{figure}

\subsection{Superposition of reconstructions}
\begin{figure}[H]
\centering
\includegraphics[width=.5 \textwidth]{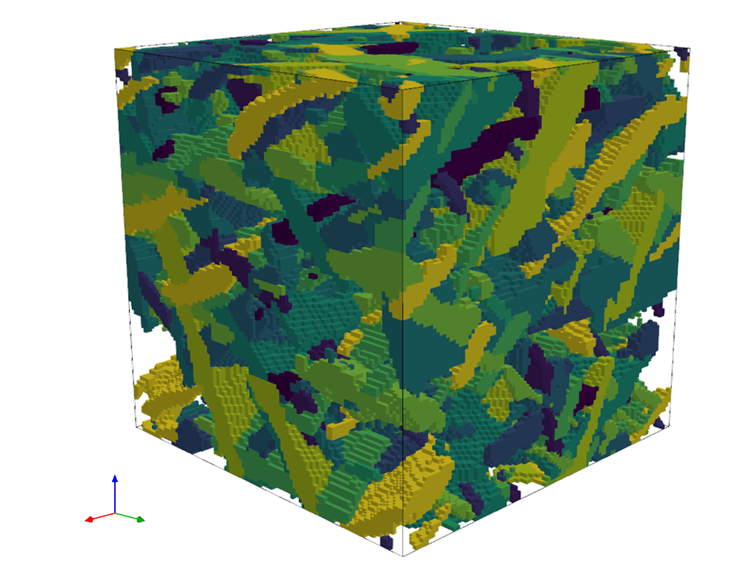}
\caption{Superposition of all twelve 3D reconstructions with a resulting volume fraction of 52.78\%}\label{rec1}
\end{figure}

The complete reconstruction of the microstructure is assembled from the reconstructions of the separate subsystems. At this stage, no specific hierarchy is governing the superposition. The resulting microstructure is depicted in Figure \ref{rec1}.

In the reconstructed microstructure, key characteristics of the titanium basketweave microstructure, like high-aspect-ratio, intersecting grains, are captured and qualitative resemblance to experimentally obtained 3D data \cite{demott20203d} is observed. A problem that becomes evident quickly is the relatively low volume fraction of filled space in the reconstruction domain. While the sum of volume fractions of the separate reconstructions is 73.11\%, the superposition of reconstructions only fills 52.78\% of the space in the reconstruction domain. The volume fraction is preserved in the 2D to 3D reconstructions, but substantial loss is observed in the superposition of reconstructions, as the separate reconstructions do not take into account the space that is to be filled by another grain group. 

\section{Conclusion}
We present a methodology for the generative reconstruction of 3D reconstructions of polycrystalline microstructures from conventional EBSD data, leveraging Burgers orientation relationship and cross-section-based optimization based on CNN-based descriptors using the MCRpy implementation. Circumventions to a few of a the limitations of the approach are proposed. A number of open challenges remain like the treatment of secondary martensite, ensuring a representative distribution of volume fractions, and quantitative evaluation of the reconstructed microstructures.

\section*{Acknowledgements}
We would like to thank Ali Reza Safi and Dr. Benjamin Klusemann from Helmholtz-Zentrum Hereon for the in-depth discussions regarding microstructure reconstruction and MCRpy usage, and Sikander Naseem from the Fraunhofer Innovation Platform for Advanced Manufacturing at the University of Twente for supplying the EBSD micrographs.

%\bibliographystyle{abbrv} 
%\bibliography{bibliography}

\end{document}